\documentstyle[12pt]{article}
\begin{document}
\title{Hara's theorem, quark model, and $\gamma _5$-dependent 
renormalization constants}
\author{
{P. \.{Z}enczykowski}$^*$\\
\\
{\em Dept. of Theor. Physics} \\
{\em Institute of Nuclear Physics}\\
{\em Radzikowskiego 152,
31-342 Krak\'ow, Poland}\\
}
\maketitle
\begin{abstract}
We discuss the applicability of $\gamma _5$-dependent field
renormalization as a means of renormalizing away
the apparent violation
of Hara's theorem observed in the quark model.
It is pointed out that a result totally analogous to
the violation of Hara's theorem is predicted  
by the quark model also for neutral baryons.
For neutral baryons, however, such a result 
cannot be renormalized away. 
This proves that $\gamma _5$-dependent 
renormalization does not provide a proper way for a hadron-level understanding
of the violation of Hara's theorem observed in the quark model.
\end{abstract}
$^*$ E-mail:zenczyko@solaris.ifj.edu.pl\\
\vskip 0.1in
\begin{center}REPORT \# 1777/PH, INP-Krak\'ow \end{center}
\newpage

Weak radiative hyperon decays (WRHD's) have been a challenge to our 
understanding for over 30 years.
Despite all the work done during these years, a
satisfactory theoretical understanding of these processes is still lacking.
For a review of current theoretical and experimental situation 
in the field see ref.\cite{LZ}.

The puzzle posed by weak radiative hyperon decays centers
on the issue of Hara's theorem \cite{Hara} and its possible violation. 
Hara's theorem
states that the parity-violating amplitude of decay
$\Sigma ^+ \rightarrow p \gamma $ should vanish in the limit of SU(3)
flavour symmetry. 
There are two results that seem to indicate that Hara's theorem is 
violated. 
\begin{enumerate}
\item 
Since SU(3) symmetry is expected to be weakly broken one expects small 
parity-violating $\Sigma ^+ \rightarrow p \gamma $
amplitude (and, consequently, 
small decay asymmetry). 
Experiment \cite{Foucher} shows, however, that the asymmetry is large:
\begin{equation}
\label{eq:Foucher}
\alpha (\Sigma ^+ \rightarrow p \gamma )= -0.72 \pm 0.086 \pm 0.045
\end{equation}
Furthermore, existing hadron-level Hara's-theorem-satisfying models
lead to a pattern of asymmetries of three related WRHD's 
($\Lambda \rightarrow n \gamma$,$\Xi ^0 \rightarrow \Lambda \gamma$,
$\Xi ^0 \rightarrow \Sigma ^0 \gamma$) that does not seem to be corroborated by
experiment.
\item 
On the theoretical side it was observed by Kamal and Riazuddin \cite{KR}
that Hara's theorem is violated in the
quark model even in the case of exact SU(3)-symmetry.
There have been several proposals of how to understand this quark model result
\cite{LL,Zen,Dmi96}.  All of them have various deficiencies.
At present there is no consensus on how the result of Kamal and Riazuddin
should be understood. The purpose of this note is to discuss the most recent
proposal \cite{Azimov} in that matter. 
\end{enumerate}
In ref.\cite{Azimov} it has been argued that the 
apparent violation of Hara's theorem obtained in the quark model 
can be renormalized away at hadronic level
by means of field renormalization with $\gamma _5$-dependent 
renormalization constants.  Here we show that such a 
renormalization procedure, although 
applicable to fundamental charged
fermions, cannot be successfully used for neutral baryons.
Since quark model predicts violation of a theorem analogous to that
of Hara also for neutral baryons, the conclusion is that 
the effect observed by Kamal and Riazuddin
should not be associated with a possible need for a
$\gamma _5$-dependent field renormalization.
Thus, a proper way of understanding
the quark model result must lie elsewhere.
 
Let us begin by writing down the most general form of 
(diagonal in flavour) vector and axial currents that couple to photon:

\begin{equation}
\label{eq:V}
V_{\mu } = \overline{\psi } [f_1(q^2)\gamma _{\mu } +
f_2(q^2) i \sigma _{\mu \nu}q^{\nu } + f_3(q^2) q_{\mu }] \psi 
\end{equation}

\begin{equation}
\label{eq:A}
A_{\mu } = \overline{\psi } [g_1(q^2)\gamma _{\mu }\gamma _{5} +
g_2(q^2) i \sigma _{\mu \nu}\gamma _5q^{\nu } + 
g_3(q^2)\gamma _5 q_{\mu }] \psi 
\end{equation}

As in ref.\cite{Azimov} the notation of \cite{BjD} is used for Dirac matrices.
CP-invariance requires reality of functions $f$ and $g$ 
if standard form of Dirac equation is used.
(In ref.\cite{Azimov}
formfactors $f_3$ and $g_2$ are considered imaginary which
is incompatible with the requirement of CP-invariance (see 
ref.\cite{Okun}).)

There are two types of $\gamma _5$-dependent transformations of Dirac
spinors:

\noindent
1) "phase" transformation
\begin{equation}
\label{eq:phase}
\psi ' = \exp{(i\alpha \gamma _5)}\;\psi
\end{equation}
2) "scale" transformation
\begin{equation}
\label{eq:scale}
\psi ' = \exp {(\beta \gamma _5)}\; \psi
\end{equation}
with real $\alpha$, $\beta$.
The adjoint spinors transform like
$\overline{\psi '} = \overline{\psi }\exp{(i\alpha \gamma _5)} $ and
$\overline{\psi '} = \overline{\psi }\exp{(-\beta \gamma _5)} $ respectively.
Spinors $\psi _L$ and $\psi _R$ transform under phase transformations 
with opposite phases while under scale transformations their relative
size is changed.

Let us discuss how the form of currents $V_{\mu}$, 
$A_{\mu}$ is affected by phase and scale transformations.
Using properties of $\gamma $ matrices we find
\begin{eqnarray}
\label{eq:vertex}
\exp{(-i\alpha \gamma _5)}[\gamma _{\mu}, \gamma _{\mu}\gamma _5 ]
\exp{(-i\alpha \gamma _5)}&=&
[\gamma _{\mu},\gamma _{\mu}\gamma _5 ] \nonumber \\
\exp{(-i\alpha \gamma _5)}
[\sigma _{\mu \nu},\sigma _{\mu \nu}\gamma _5,\gamma _5]
\exp{(-i\alpha \gamma _5)}&=&
[\sigma _{\mu \nu},\sigma _{\mu \nu}\gamma _5,\gamma _5]
(c-is\gamma _5) \nonumber \\
\exp(\beta \gamma _5)
[\sigma _{\mu \nu},\sigma _{\mu \nu}\gamma _5,\gamma _5]
\exp(-\beta \gamma _5)&=&
[\sigma _{\mu \nu},\sigma _{\mu \nu}\gamma _5,\gamma _5]\nonumber \\
\exp(\beta \gamma _5)[\gamma _{\mu},\gamma _{\mu} \gamma _5] 
\exp(-\beta \gamma _5)&=&
[\gamma _{\mu},\gamma _{\mu} \gamma _5]
(c_h-s_h\gamma _5) 
\end{eqnarray}
where $c=\cos 2\alpha$, $s= \sin 2\alpha$,
$c_h=\cosh 2\beta$, $s_h=\sinh 2\beta $.

Under phase transformations
the standard form of currents (Eqs.(\ref{eq:V},\ref{eq:A})) 
transforms therefore to
\begin{equation}
\label{eq:V'}
V'_{\mu } = \overline{\psi '} [f_1\gamma _{\mu } +
(c f_2-i s g_2)i \sigma _{\mu \nu}q^{\nu } + 
(c f_3-i s g_3) q_{\mu }] \psi '
\end{equation}
\begin{equation}
\label{eq:A'}
A'_{\mu } = \overline{\psi '} [g_1\gamma _{\mu }\gamma _{5} +
(c g_2-i s f_2) 
i \sigma _{\mu \nu}\gamma _5q^{\nu } + 
(c g_3-i s f_3)\gamma _5 q_{\mu }] \psi '
\end{equation}
From Eqs.(\ref{eq:V'},\ref{eq:A'})
we see that: 
1) functions $f'_1$ $(=f_1)$, $g'_1$ $(=g_1)$ are unaffected
by phase transformations and
2) functions $f'_2$ $(=cf_2-isg_2)$, $f'_3$, $g'_2$, $g'_3$ 
may in general be complex
even though CP is conserved.
However, the form of the Dirac equation has to be modified then to
(compare ref.\cite{DGH})
\begin{equation}
\label{eq:massterm}
(p\!\!/-m(c-is\gamma _5))\psi ' = 0
\end{equation}

Similarly, 
under scale transformations the standard form (Eq.(\ref{eq:V},\ref{eq:A}))
of currents $V_{\mu}$, $A_{\mu}$ 
transforms to
\begin{equation}
\label{eq:V''}
V'_{\mu } = \overline{\psi '} [f_1'\gamma _{\mu } +
f_2 i \sigma _{\mu \nu}q^{\nu } + f_3 q_{\mu }] \psi '
\end{equation}
\begin{equation}
\label{eq:A''}
A'_{\mu } = \overline{\psi '} [g_1'\gamma _{\mu }\gamma _{5} +
g_2 i \sigma _{\mu \nu}\gamma _5q^{\nu } + 
g_3\gamma _5 q_{\mu }] \psi '
\end{equation}
where $f_1'= c_h f_1 - s_h g_1$, $g_1'= c_h g_1 - s_h f_1$, and $f_2=f'_2$ etc.
i.e. only the coefficients at the 
$\gamma _{\mu}$, $\gamma _{\mu}\gamma _5$ terms 
are modified.
Hereafter the prime sign ($'$) is used to label functions $f$, $g$ when they
correspond to a non-standard form of Dirac equation.

We are now prepared to discuss the applicability of 
$\gamma _5$-dependent renormalization to neutral baryons.
We follow the argument of ref.\cite{Azimov} closely.
For simplicity consider just a neutron.
Near its mass-shell and
in the absence of weak interactions (but with complete
account for strong and electromagnetic interactions) 
the neutron propagator has the form
\begin{equation}
\label{eq:bare}
S^{-1}_o = p\!\!/-m_o
\end{equation}
Let us now set the Cabibbo angle to zero.
With weak interactions turned on the propagator of Eq.(\ref{eq:bare}) 
is modified and close to
its new mass-shell it has the general form (see ref.\cite{Azimov})
\begin{equation}
\label{eq:S'}
S'^{-1}= ap\!\!/+bp\!\!/\gamma _5 - m'
\end{equation}
In writing Eq.(\ref{eq:S'}) we have assumed that an appropriate phase
transformation has been already carried out to bring the general mass
term of the form given in Eq.(\ref{eq:massterm}) to the standard form.
Such a transformation does not affect the $ap\!\!/$ and 
$bp\!\!/ \gamma _5$ terms (compare Eq.(\ref{eq:vertex})).
Since weak interactions are small perturbations we have $a \approx 1$,
$|b/a| \ll 1$, and $m' \approx m_o$.
Let us now bring the propagator of Eq.(\ref{eq:S'}) to the standard
Dirac form.
In order to achieve this we perform a scale renormalization with appropriate
parameter $\beta$:
\begin{eqnarray}
\label{eq:S}
S^{-1}&=&\exp(-\beta \gamma _5)S'^{-1}\exp(+\beta \gamma _5) \nonumber \\
       &=& (a\, c_h + b\, s_h)p\!\!/ + (b \,c_h + a \,s_h) 
       p\!\!/\gamma _5 - m'
\end{eqnarray}
Thus, in order to bring neutron propagator to its standard Dirac form we 
need $\tanh (2\beta) = - b/a$, i.e. $\beta $ of order $b/a$.

Renormalization of propagators as in Eq.(\ref{eq:S}) is associated with a
simultaneous renormalization of fields ($\psi ' \rightarrow \psi$, as in 
Eq.(\ref{eq:scale}))
and of the form of currents 
(Eqs.(\ref{eq:V''},\ref{eq:A''})).
Since we require that, after renormalization, neutron couplings to photon 
satisfy
$f_1(0) = 0$ (zero charge) and $g_1(0)=0$ (analog of the
assumption necessary for the proof of Hara's theorem)
we obtain the conditions
\begin{eqnarray}
\label{eq:conditions}
f_1(0)=f'_1(0) c_h + g'_1(0) s_h &=& 0 \nonumber \\
g_1(0)=g'_1(0) c_h + f'_1(0) s_h &=& 0
\end{eqnarray}
Assume now that in perturbative calculations in
some model we have obtained a nonvanishing $g'_1(0)$. 
Whatever value of $f'_1(0)$ is obtained it is clear 
that renormalization conditions of
Eqs.(\ref{eq:conditions}) require $f'_1(0)/g'_1(0)= 
-\tanh(2\beta)= -1/\tanh(2\beta)$.
Thus, in particular, $\beta = \pm \infty$ is required.
This cannot be reconciled with the perturbative renormalization condition
that $\beta$ is to be of order $|b/a|\ll 1$. Thus, if a nonzero $g'_1(0)$
is somehow generated for {\em neutral} baryons, it \underline{cannot} 
be renormalized away by a $\gamma _5$-dependent transformation.

Let us now show that in the quark model the perturbative calculation 
of the contribution from $W$-exchange between quarks 
does indeed lead to $g'_1(0) \ne 0$ for neutron $n$. 
To see this observe that neutron spin-flavour wave function (quarks: $ddu$) 
is obtained from that of proton (quarks: $uud$) by
a simple replacement $u \leftrightarrow d$.
Symmetry of the wave function ensures that it is sufficient to consider
$W$-exchange in one $ud$ diquark only.  
For the proton the photon-proton parity-violating 
coupling can be expressed in terms of
photon-diquark couplings as
\begin{equation}
\label{eq:proton}
\langle p \uparrow \gamma | T | p \downarrow \rangle =
\frac{1}{3\sqrt{2}}t_{+1}-\frac{1}{3\sqrt{2}}t_{-1}-\frac{1}{\sqrt{2}}v
\end{equation}
where parity violating
weak+electromagnetic {\em diquark} $\rightarrow$ {\em diquark} + $\gamma$
transition amplitudes are defined as
\begin{eqnarray}
\label{eq:diquark}
t_{+1}&=&\langle S^{+1}(ud)\gamma | T | A(ud)\rangle\nonumber \\
t_{-1}&=&\langle A(ud)\gamma | T | S^{-1}(ud)\rangle\nonumber \\
v&=&\langle S^0(ud) \gamma | T | S ^{-1}(ud) \rangle +
\langle S^{+1}(ud) \gamma | T | S ^0(ud) \rangle
\end{eqnarray}
with diquark states
\begin{eqnarray}
\label{eq:diquarks}
|A(ud)\rangle &=&
|\frac{1}{\sqrt{2}}(ud-du)
\frac{1}{\sqrt{2}}(\uparrow \downarrow-\downarrow \uparrow)\rangle \nonumber \\
|S^{+1}(ud)\rangle &=& 
|\frac{1}{\sqrt{2}}(ud+du) \uparrow \uparrow \rangle
\nonumber \\
|S^{0}(ud)\rangle &=& 
|\frac{1}{\sqrt{2}}(ud+du)
\frac{1}{\sqrt{2}}(\uparrow \downarrow+\downarrow \uparrow)\rangle \nonumber \\
|S^{-1}(ud)\rangle &=&
|\frac{1}{\sqrt{2}}(ud+du) \downarrow \downarrow \rangle
\end{eqnarray}
For the neutron we get similarly
\begin{equation}
\label{eq:neutron}
\langle n \uparrow \gamma | T | n \downarrow \rangle =
-\frac{1}{3\sqrt{2}}t_{+1}+\frac{1}{3\sqrt{2}}t_{-1}-\frac{1}{\sqrt{2}}v
\end{equation}
Calculations in the quark model as in ref.\cite{KR} give
$t_{+1}=-t_{-1}$ and $v=0$ and, consequently, they yield equal (up to a sign) 
nonzero values of $g'_1(0)$ for proton and neutron.
The origin of the nonzero value of $g'_1(0)$ (whatever it is)
is clearly the same for 
both proton and neutron.

The above argument may be dressed in a slightly more elaborate form by
following the lines of ref.\cite{Azimov}.  Namely, one can consider
a Cabibbo-suppressed charm-changing weak radiative transition $\Sigma ^0_c
\rightarrow n \gamma$ ($cdd \rightarrow udd + \gamma$).
For this process an analog of Hara's theorem is expected to hold: in the limit
of equal masses of the $u$ and $c$ quarks the parity violating $\Sigma ^0_c
\rightarrow n \gamma$
amplitude should vanish.
The reasoning of ref.\cite{Azimov} may then be applied. 
First, we imagine a world in which $u$ and $c$ quarks are degenerate.
By forming appropriate linear combinations of $u$ and $c$ quarks ($u'$,$c'$) 
one can
then eliminate the $c' \rightarrow d$ transition in the Cabibbo matrix
and deal with weak $W$-exchange-induced 
$u'd \rightarrow du'$ transition exclusively (and a
vanishing $c'd \rightarrow dc'$ transition).
States $n'$ and $\Sigma '^o_c$ are then not transformed into each
other by a single $W$-exchange process.  
In this way we diagonalize the problem and are led
to consider process $u'dd \rightarrow u'dd +\gamma$
i.e. an $n' \rightarrow n' \gamma$ coupling.  This is essentially
what we were discussing previously with an unimportant change of names
($n' \rightarrow n$).  

Inability of the $\gamma _5$-dependent 
field renormalization to explain the origin of quark-model
violation of Hara's theorem is clearly related to
the composite nature of baryons as prescribed by the quark model.
Indeed, the $W$-exchange contribution which lies at the origin of
quark-model violation of Hara's theorem "does not know" about the
charge of the spectator quark. Thus the total baryon
charge may be reduced to zero by assuming
an appropriate charge on the spectator quark. 
Discussion of this paper shows therefore that if
the result of Kamal and Riazuddin is to be understood at some composite level
it should probably be the diquark level.

In summary
we have shown that:
\noindent
1) quark model does lead to a nonvanishing $g'_1(0)$
for neutron-photon couplings and 
\noindent
2) this contribution cannot be renormalized
away by a $\gamma _5$-dependent transformation. 

Consequently, 
$\gamma _5$-dependent renormalization does not provide a proper way for 
a hadron-level
understanding of the violation of Hara's theorem in the quark model.

ACKNOWLEDGEMENTS.

This work was partially supported by the KBN grant No 2P03B23108.

\end{document}